\documentclass[twocolumn]{article}

\usepackage{amsmath,amssymb,amsfonts}
\usepackage{algorithm}
\usepackage{algorithmic}
\usepackage{graphicx}
\usepackage{subfig}
\usepackage{textcomp}
\usepackage{array}

\usepackage{dsfont}
\usepackage{multirow}
\usepackage{fancyhdr}
\usepackage{wrapfig}
\usepackage[T1]{fontenc}
\usepackage[utf8]{inputenc}
\usepackage{authblk}
\usepackage{color}
\usepackage{xcolor}
\newcommand{\hl}[1]{\textcolor{black}{{#1}}}
\newcommand{\BTT}[1]{\textcolor{black}{{#1}}}

\title{Challenging the adversarial robustness of DNNs based on error-correcting output codes}
\author[a]{Bowen Zhang}
\author[b]{Benedetta Tondi}
\author[a\,*]{Xixiang Lv}
\author[b]{Mauro Barni}
\affil[a]{School of Cyber Engineering, Xidian University, 710126 China}
\affil[b]{Department of Information Engineering and
	Mathematics, University of Siena, 53100, Italy}
\affil[*]{Correspondence should be addressed to Xixiang Lv; xxlv@mail.xidian.edu.cn}

\begin{document}
\maketitle
\begin{abstract}
The existence of adversarial examples and the easiness with which they can be generated raise several security concerns with regard to deep learning systems, pushing researchers to develop suitable defense mechanisms.
The use of networks adopting error-correcting output codes (ECOC) has recently been proposed to counter the creation of adversarial examples in a white-box setting.
In this paper, we carry out an in-depth investigation of the adversarial robustness achieved by the ECOC approach. We do so by proposing a new adversarial attack specifically designed for multi-label classification architectures, like the ECOC-based one, and by applying two existing attacks.
In contrast to previous findings, our analysis reveals that ECOC-based networks can be attacked quite easily by introducing a small adversarial perturbation. Moreover, the adversarial examples can be generated in such a way to achieve high probabilities for the predicted target class, hence making it difficult to use the prediction confidence to detect them. %
\hl{Our findings are proven by means of experimental results obtained on MNIST, CIFAR-10 and GTSRB classification tasks.}
\end{abstract}

\section{Introduction}

\label{sec:introduction}
Deep neural networks can solve complicated computer vision tasks with unprecedented high accuracies.
However, they have been shown to be vulnerable to \textit{adversarial examples}, namely, properly crafted inputs introducing small (often imperceptible) perturbations, inducing a classification error \cite{Intriguing, Explaining and Harnessing, Access Survey}. The possibility of crafting both non-targeted and targeted attacks have been demonstrated, the goal of the former being to induce any kind of classification error \cite{deepfool,UniversalPerturbation}, while the latter aim at making the network decide for a target class chosen a priori \cite{Intriguing,CW attack}. \hl{It goes without saying that, in general, targeted attacks are more difficult to build.}

As a reaction to the threats posed by adversarial examples, many defense mechanisms have been proposed to increase the adversarial robustness of deep neural networks \cite{Defense1, Defense2, Defense3,EnsembleAdvTraining,advtraining1,advtraining2}. However, in a white-box setting wherein the attacker has a full knowledge of the attacked network, including full knowledge of the defense mechanism, more powerful attacks can be developed, thus tipping again the scale in favor of the attacker \cite{DefenseIsNotEasy,UniversalPerturbation}.

In this race of arms, a novel defence strategy based on \textit{Error-Correcting Output Coding} (ECOC) \cite{ori ECOC}
 has been proposed recently in \cite{ECOC}, to counter adversarial attacks in a white-box setting.
More specifically, given a general multi-class classification problem, error-correcting output codes are used to encode the various classes and represent the network's outputs.
To explain how, let us refer to the output of the last layer of the network, prior to the final activation layer, as logit values or simply {\em logits}.
In general, the final activation layer consists of the application of an activation function, that maps the logits into a prescribed range, and a normalization layer, that maps the output of the activation functions into a probability vector, associating a probability value to each class. In the common case of one-hot-encoding, a softmax layer is used, in which case these two steps are performed simultaneously. During training, the network learns to output a large logit value for the true class and small values for all the others. With the ECOC approach, instead, the network is trained in such a way to produce normalized logit values that correlate well with the codeword used to encode the class the input sample belongs to. In general, ECOC codewords have many non-zero values thus marking a significant difference with respect to the one-hot-encoding case.

The rationale behind the use of the ECOC architecture to counter the construction of adversarial examples \cite{ECOC}, is that while with classifiers based on standard one-hot-encoding the attacker can induce an error by modifying one single logit (reducing the one associated to the ground-truth class or increasing the one associated to the target class), the final decision of the ECOC classifier depends on multiple logits in a complicated manner, and hence it is supposedly more difficult to attack, (especially when longer codewords are used).

In \cite{ECOC}, the authors considered non-targeted attacks in their experiments, and showed with the popular white-box C\&W attack that the attack success rate on CIFAR-10 \cite{cifar10}, passes from 100\% - for one-hot-encoding - to 29\% - for an ECOC-based classifier. 

Another alleged advantage of the ECOC architecture proposed in \cite{ECOC} is linked to the way the probabilities associated to each class are computed. Rather than using a softmax function as commonly done with one-hot-encoding, first the correlation between the activated outputs and the codeword is computed, then a linear normalization procedure is applied (see eq. (\ref{eq:ecoc_map}) in the following).
In this way, the probability assigned to the class chosen by the classifier grows more smoothly and samples close to the decision region boundary (like adversarial examples are likely to be) are classified with a low confidence. Results presented in \cite{ECOC}, in fact,  show that the ECOC model tends to provide sharp results for clean images, while it is often uncertain about the (incorrect) prediction made on adversarial examples. This behavior could be exploited to, at least, distinguish between adversarial examples and benign inputs.

The goal of this paper is to further verify if and to which extent the use of error correction codes to encode the output of deep neural networks allows to increase the robustness against targeted adversarial examples. We do so by introducing a new white-box attack, inspired to C\&W attack, explicitly thought to work not only ECOC against, but also other multi-label classifiers. In fact, the original C\&W is naturally designed to deceive networks adopting the one-hot-encoding strategy, and it looses some of its advantages when used against ECOC systems. \hl{We stress that, in contrast to previous works (see, for instance, \cite{ECOC} and Section 10 in \cite{AdaptiveAttack}), we aim at developing a targeted attack, which is a more difficult task than crafting non-targeted adversarial examples. This is a reasonable choice for at least two reasons. First, targeted attacks are more flexible than non-targeted ones, since they can be used in a wider variety of applications, wherein the ultimate goal of the attack may vary considerably. Secondly, being able to attack a defence under most stringent attacking constraints illustrates better the weakness of the defence itself.}

\hl{We run extensive experiments to evaluate the ability of ECOC-based classifiers to resist the new attack and compared the results we got with those obtained by applying a fine-tuned version of C\&W attack and the LOTS attack introduced in \cite{LOTS}. The experiments were carried out by considering three different classification tasks, namely traffic sign classification (GTSRB)\cite{GTSRB}, CIFAR-10 classification\cite{cifar10} and MNIST\cite{MNIST}. As a result, we found  that the ECOC classifiers can be successfully attacked with high success rate.
In particular, the new attack outperforms the other two especially when long codewords are used by ECOC.
We also verified that, by increasing the confidence of the attack, adversarial examples can achieve high probabilities for the predicted target class, similar to those of benign samples, hence making it difficult to use the prediction confidence to detect adversarial samples.
Overall, our analysis reveals that the security gain achieved by the ECOC scheme is a minor one, thus calling for more powerful defences.}

The rest of this paper is organised as follows, we first briefly review the ECOC scheme presented in \cite{ECOC}, then we describe the proposed attack. The setup considered for the experiments and the results we got are reported and discussed in the Experiments section. Eventually, we review the related work at the end of the paper.

\section{ECOC-based classification}
\begin{figure*}
	\centering{\includegraphics[width =0.95\textwidth]{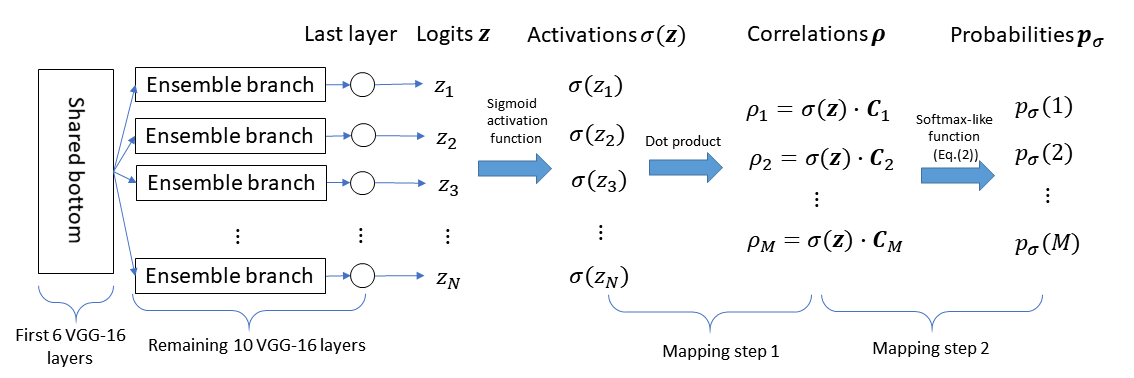}}
	\caption{Block diagram of ECOC architecture.}
	\label{fig:ecoc}
\end{figure*}
Let us first introduce the notation for a general multi-class CNN.
Let $x$
be the input of the network and $k$ the class label, $k=1,2...,M$, where $M$ denotes the number of classes. Let  $f(x)$ indicate the decision function of the network. We denote by $\mathbf{z} = (z_1,z_2,...)$, the  vector with the logit values, that is, the network values before the final activations and the mapping to class probabilities. For one-hot-encoding schemes, $\mathbf{z}$ has length $M$ and the logits are directly mapped into probability values through the
softmax function $\psi$ as follows:

\begin{equation}\label{eq:softmax}
p_{\psi}(k) = \psi_{k}({\bf z}) =  \frac{\exp(z_k)}{\sum_{i=1}^{M}\exp(z_i)},
\end{equation}
for $k = 1,..,M$.
Then, the final prediction is made by letting $f(x) = {\arg\max}_{k} \hspace{0.1cm} p_{\psi} (k)$.

The  error-correction-output-coding (ECOC) scheme proposed in \cite{ECOC}  assigns a codeword $\mathbf{C}_k$ of length $N$ ($N \geq M$) to every output class ($k = 1,...,M$).  $\mathbf{C}$ denotes the  $M \times N$ codeword matrix.  Each element of $\mathbf{C}$ can take values in  $\{-1,1\}$.
In this way, the length of the logit vector $\mathbf{z}$ is $N$.
The logits are first mapped into the $[-1, 1]$ range by means of an activation function $\sigma(\cdot)$ \BTT{(e.g. the tanh function that $ \sigma(x) = (e^x-e^{-x}) / (e^x + e^{-x}) $)}. Then, the probability of class $k$ is computed by looking at the correlation with $\mathbf{C}_k$, according to the following formula:
\begin{equation}\label{eq:ecoc_map}
p_{\sigma}(k) = \frac{\max(\sigma(\mathbf{z})\cdot \mathbf{C}_k, 0)}{\sum_{i=1}^{M} \max(\sigma(\mathbf{z})\cdot \mathbf{C}_i,0)}
\end{equation}
where $\cdot$ denotes the inner product and $\sigma(\cdot)$ is a sigmoid activation function applied element-wise to the logits. Since $C_{ij}$'s take values in  $\{-1,1\}$, the {\em max} is necessary to avoid negative probabilities. \BTT{According to \cite{ECOC}, the common softmax rule (Eq.\eqref{eq:softmax})) is able to express uncertainty between two classes only when the logits are roughly equal (i.e. $z_1 \approx z_2  $ and the two probabilities are close $ p_\psi (i) \approx p_\psi (j) $). In a two dimensional case, this corresponds to a very narrow stripe, approximate to a line, across the boundary of the decision region, while in high dimensional spaces, the region $ z_i \approx z_j $ approximates a hyperplan, a  $ M-1 $ dimensional subspace of $ \mathbb{R}^M $ with negligible volume, and hence the classifier outputs high probabilities almost everywhere.} This makes it very easy for the attacker to find an adversarial input that is predicted (wrongly) with high confidence. \BTT{With ECOC (Eq.\eqref{eq:ecoc_map}), instead, it is sufficient that two approximate correlations express low uncertainty ($ \sigma(\mathbf{z})\cdot \mathbf{C}_i \approx \sigma(\mathbf{z})\cdot \mathbf{C}_j$), then a non-trival volume is allocated to low-confidence region in the logit space, thus limiting the freedom of the attacker to craft high-confidence adversarial examples.}
An overall sketch of the ECOC scheme is depicted in Figure \ref{fig:ecoc}.
The logits $\mathbf{z}$  are first mapped into correlation values, ${\boldsymbol{\rho}}:= \sigma(\mathbf{z})\cdot \mathbf{C}$ (mapping step 1), then the vector with the correlations is normalized so to form a probability distribution (mapping step 2) via the normalization function in \eqref{eq:ecoc_map}.
The model's final predicted label is 
${\arg\max}_{k} \hspace{0.1cm} p_\sigma (k)$.
Eq. \eqref{eq:ecoc_map} is a generalization of the standard softmax activation in eq. \eqref{eq:softmax}
and reduces to it for the case of one-hot-encoding, that is when $\mathbf{C} = \mathbf{I}_{M}$, with $N = M$, and where $\mathbf{I}_{M}$ is the identity $M \times M$ matrix.

The purpose of the ECOC architecture is to design a classifier which is robust to changes of multiple logits, and then, expectedly, more difficult to attack (with standard one-hot-encoding the adversary can succeed by altering a single logit). For the scheme to be effective, codewords characterised by a large minimum Hamming distance must be chosen.
For simplicity, in \cite{ECOC}, the ECOC classifier is built by using Hadamard codes taking values in $\{-1,1\}$ (when $\mathbf{C}$ is a Hadamard matrix, the Hamming distance for large $M$ approaches the limit value $N/2$).  An advantage with this choice is that, since  $\mathbf{C}$ is orthogonal, whenever the network outputs a codeword exactly (that is when $\sigma(\bf z) = C_k$), then $p_{\sigma}(k) = 1$.
The tanh function is selected as the activation function $\sigma(\cdot)$.

The authors also found that, rather than considering a single network with $N$ outputs, a classifier consisting of an ensemble of several smaller networks, each one outputting a few codeword elements, permits to achieve a larger robustness against attacks. By training separate networks, in fact, the correlation between errors affecting different bits of the codewords is reduced, thus forcing the attacker to attack all the bits independently.
In the scheme in Figure \ref{fig:ecoc}, every network outputs one codeword bit only, resulting in $N$ ensemble branches.
\section{Attacking ECOC}\label{sec:attack}
We start by considering the basic C\&W attack introduced in \cite{CW attack}. We notice that some of the good properties of C\&W do not longer hold when the attack is applied against the ECOC scheme, since it has been originally designed to work against networks adopting the one-hot-encoding strategy. Then we propose a new more effective attack, which is specially tailored to multi-label structures like ECOC.

\BTT{In general, constructing an adversarial example corresponds to finding a small perturbation $ \delta $ (under some distance metric) that once added to image $ x $ will change its classification. Such a problem is usually formalised as:}
\begin{equation}
\begin{aligned}
	\min \quad & \mathcal{D}(x, x+\delta)\\
	\textrm{s.t.} \quad &  f(x+\delta)  = t\\
\end{aligned}
\end{equation}
\BTT{where $ \mathcal{D} $ is some distance metric (e.g. the $L_2$ metric) and $ t $ is a chosen target class. As this problem is difficult to solve, C\&W attack aims at solving its Lagrangian approximation, defined as:}
\begin{equation}\label{eq:CW one-hot}
\min ||\delta||_2 + \lambda\cdot\max( \underset{i \neq t}{\max}(z_i(x+\delta)) - z_t(x+ \delta), c)
\end{equation}
\BTT{where the second term is any function such that $ f(x+\delta) = t $ if and only if this term $  \leq c$. $||\cdot||_2$ denotes the $L_2$-norm, $\lambda$ and $c$ are constant parameters ruling, respectively, the tradeoff between the two terms of the optimization problem, and the confidence margin of the attack\footnote{\scriptsize{\BTT{In \cite{CW attack}, $\delta = 1/2 (\tanh(w) + 1)-x$ and the minimization is carried out over $w$ to have box constraints on $ \delta $ when optimizing Eq.\eqref{eq:CW one-hot} with a common optimizer like Adam.}}}. Eq.\eqref{eq:CW one-hot} is designed for the common one-hot encoding case. In fact, it is easy to see that for ECOC the motivation of such a design does not hold anymore, and ensuring that the second term is less than $ c $ does not guarantee that $ f(x+\delta)=t $. Therefore the C\&W attack must be adjusted to fit the ECOC framework. By noting that in ECOC correlations are proportional to probabilities (instead of the logits as with one-hot encoding), C\&W shall be modified as:}
\begin{equation}\label{eq:CWC}
\underset{\mathbf{\delta}}{\text{minimize}}\; ||\mathbf{\delta}||_2 + \lambda \cdot \max( \underset{i \neq t}{\max}(\rho_i(x+\delta)) - \rho_t(x+ \delta), c),
\end{equation}
\BTT{where $ \rho_t(x+\delta) =  \sigma(\mathbf{z}(x+\delta))\cdot \mathbf{C}_t$.}.

A key advantage of C\&W attack against one-hot-encoding networks is that it works directly at the logits level. In fact, logits are more sensitive to modifications of the input than the probability distribution obtained after the softmax activation\footnote{\scriptsize{Most adversarial attacks work directly on the probability values obtained after the softmax, which makes them less effective than C\&W, and prone to gradient-vanishing problems.}}.

When C\&W attack is applied against ECOC (by means of \eqref{eq:CWC}), it does not work at the output logit level, but after that the correlations are computed (mapping step 1), since this is the layer that precedes the application of the softmax-like function.
The correlations between the activations of the logits and the codewords will likely have a reduced sensitivity to input modifications,
and this may decrease the effectiveness of the attack. 
We also found that during the attack, it is possible to change only one bit of the output while the others are almost unchanged. This can be explained by observing that ECOC trains each output bit separately, so that each bit can be treated as an individual label. In this way the correlation between the output bits is significantly reduced compared to classifiers adopting the one-hot encoding approach. We exploit this fact to design our attack in such a way as to make it modify a single bit at a time, and iteratively repeat this process to eventually change multiple bits.

With the above ideas in mind, the new attack is formulated as follows:
\begin{equation}\label{eq:CWB}
\underset{\mathbf{\delta}}{\text{minimize}}\; ||\mathbf{\delta}||_2 - \lambda \cdot \underset{i}{\min}( 2 t_i \cdot z_i(x + \delta), c),
\end{equation}
where $(t_1, t_2, \cdots, t_N ) = {\bf C}_t$ is the desired target codeword ($t_i \in \{-1, 1\}$),  $\lambda $ is a parameter controlling the tradeoff between the two terms of the objective function, and $ c $ is a constant parameter used to set a confidence threshold for the attack. Specifically, the attack seeks to minimize \eqref{eq:CWB} until the product between $ t_i $ and $ z_i $ reaches this threshold, thus a higher $ c $ will result in adversarial examples exhibiting a higher correlation with the target codeword, that is adversarial examples that are (wrongly) classified with a higher confidence.

The choice of $ \lambda $ also plays an important role in the attack, given that a very small $ \lambda $ would lead to a vanishing perturbation. On the contrary, using a larger $ \lambda $ results in a more effective attack at the cost of a larger perturbation. To optimize the value of $\lambda$, we use a binary search similar to the one used in \cite{CW attack} to determine the optimum value of $ \lambda $ in C\&W attack. By doing so, the parameters of the proposed attack have the same meaning of those in C\&W attack, thus the two methods can be compared on a fair basis under the same parameter setting. 
An overall description of our attack is given in algorithm \ref{OP algorithm}, whose goal is to find a valid adversarial example, with the desired confidence level $c$ and with the smallest perturbation. As a result of the optimization in algorithm \ref{OP algorithm}, all logit value $ z_i $ of the resulting adversarial image will tend to be highly correlated with $ t_i$. 

\begin{algorithm}[h!]
	\caption{Solving optimization in \eqref{eq:CWB}}
	\label{OP algorithm}
	\begin{algorithmic}[1]
		\REQUIRE ~~\\ 
		The start point and number of binary search $ \lambda_1 $, $ n $;\\
		The step size and max iteration of gradient descent, $\epsilon$, $ m $;\\
		To be attacked image, $x$;
		\ENSURE ~~\\ 
		Adversarial perturbation $ \delta $
		\STATE $ upper\,bound \leftarrow \infty$
		\STATE $ lower\,bound \leftarrow 0 $

		\FOR{ $i \in [1,n]$}
		\STATE $ \delta_1 \leftarrow 0 $
		\STATE  $found\,adv \leftarrow $ False
		\FOR{$ j \in [1, m] $}
		\IF {$x + \delta_j $ is adversarial and $ ||\delta_j||<||\delta|| $ }
		\STATE $found\,adv \leftarrow $ Ture
		\ENDIF
		\STATE $ \delta_j \leftarrow \delta_j - \epsilon \times \frac{\nabla_i}{||\nabla_i||} $,
		where $ \nabla_i $ is the gradient of eq.\eqref{eq:CWB} with current $ \lambda_i $ w.r.t. the perturbation $ \delta_j $
		\ENDFOR
		\IF{$ found\,adv = $ True}
		\STATE $ upper\,bound  \leftarrow \lambda_i $
		\ELSE
		\STATE $ lower\,bound  \leftarrow \lambda_i$
		\ENDIF
		\IF {$ upper\,bound= \infty$}
		\STATE $ \lambda_i \leftarrow 10 \times \lambda_i $
		\ELSE
		\STATE $ \lambda_i \leftarrow (upper bound + lower bound) / 2 $
		\ENDIF
		\ENDFOR
	\end{algorithmic}
\end{algorithm}

\hl{It is worth observing that, even if we designed the new attack explicitly targeting the ECOC classifier, the algorithm in \eqref{eq:CWB} is generally applicable to any multi-label classification network since it manipulates the output bits of the network, regardless of the adopted coding strategy. This point can be evidenced by considering two limit cases of ECOC. In the first case, we avoid using error correction to encode the output classes. This is equivalent to multi-label classification problems with $N$ labels \cite{multi-label1,multi-label2}, and the proposed attack can still be applied. In the second case, we may consider one-hot-encoding as a particular way of encoding the output class. This perspective, that has also been considered in \cite{ECOC}, would degrade the ECOC system to a common network that uses one-hot-encoding and softmax to solve a multi-class classification task. Since our attack does not involve the decoding part of the network, it can still be applied to such networks.}

\section{Experiments}

\begin{table}[!h]

	\caption{Results of the attack against ECOC for GTSRB classification.
		Reported parameters indicate respectively: (start point, number of steps of binary search, max iterations, confidence).
	}
		\label{tab:GTSRB-1}
		\scalebox{0.7}{	
		\begin{tabular}{m{1.6cm}|l|l|l|l|l|l|}
			\cline{2-7}
			& \multicolumn{2}{l|}{Proposed (ECOC)} & \multicolumn{2}{l|}{C\&W (ECOC)} & \multicolumn{2}{l|}{C\&W (one-hot)} \\ \hline
			\multicolumn{1}{|m{1.6cm}|}{Parameters}       & ASR              & PSNR             & ASR            & PSNR           & ASR             & PSNR             \\ \hline
			\multicolumn{1}{|m{1.6cm}|}{(1e-4,5, 100,0)}   & 40.3\%             & 41.43            & 7.0\%            & 48.80          & 25.5\%            & 46.82            \\ \hline
			\multicolumn{1}{|m{1.6cm}|}{(1e-4,5, 200,0)}   & 45.6\%             & 41.58            & 8.6\%           & 48.48          & 37.5\%            & 45.73            \\ \hline
			\multicolumn{1}{|m{1.6cm}|}{(1e-4,5, 500,0)}   & 53.3\%             & 41.65            & 11.3\%           & 48.03          & 47.5\%            & 46.07            \\ \hline
			\multicolumn{1}{|m{1.6cm}|}{(1e-2,5, 500,0)}   & 70.6\%             & 39.85            & 20.0\%           & 43.94          & 61\%            & 43.41            \\ \hline
			\multicolumn{1}{|m{1.6cm}|}{(1e-1,10, 2000,0)} & 93.3\%             & 38.97            & 42.3\%           & 39.20          & 81.5\%            & 42.51            \\ \hline
		\end{tabular}
		}
\end{table}

\begin{table}[h]
	\caption{Results of the attack against ECOC for CIFAR-10 classification.
		Reported parameters indicate respectively: (start point, number of steps of binary search, max iterations, confidence). 
	}
	
		\label{tab:Cifar-1}
		\scalebox{0.7}{	
		\begin{tabular}{m{1.6cm}|l|l|l|l|l|l|}
			\cline{2-7}
			& \multicolumn{2}{l|}{Proposed (ECOC)} & \multicolumn{2}{l|}{C\&W (ECOC)} & \multicolumn{2}{|l|}{C\&W (one-hot)} \\ \hline
			\multicolumn{1}{|m{1.6cm}|}{Parameters}       & ASR              & PSNR             & ASR            & PSNR           & ASR              & PSNR            \\ \hline
			\multicolumn{1}{|m{1.6cm}|}{(1e-4,5, 100,0)}   & 69.3\%             & 38.59            & 53.6\%           & 39.71          & 92.5\%             & 40.03           \\ \hline
			\multicolumn{1}{|m{1.6cm}|}{(1e-4,5, 500,0)}   & 88.0\%             & 38.52            & 62.3\%           & 39.94          & 100\%            & 40.27           \\ \hline
			\multicolumn{1}{|m{1.6cm}|}{(1e-4,10, 200,0)}  & 90.6\%             & 37.84            & 79\%           & 37.32          & 100\%            & 40.18           \\ \hline
			\multicolumn{1}{|m{1.6cm}|}{(1e-4,10, 500,0)}  & 95.0\%             & 38.39            & 82.6\%           & 37.55          & 100\%            & 40.30           \\ \hline
			\multicolumn{1}{|m{1.6cm}|}{(1e-1,10, 2000,0)} & 98.6\%             & 38.41            & 92.6\%           & 36.97          & 100\%            & 39.99           \\ \hline
		\end{tabular}
		}
\end{table}

\begin{table}[!h]
	\caption{\hl{Results of the attack against ECOC for MNIST classification.
		Reported parameters indicate respectively: (start point, number of steps of binary search, max iterations, confidence)}.
	}
		\scalebox{0.7}{	
		\label{tab:MNIST}	
		\begin{tabular}{m{1.6cm}|l|l|l|l|l|l|}
			\cline{2-7}
			& \multicolumn{2}{l|}{Proposed (ECOC)} & \multicolumn{2}{l|}{C\&W (ECOC)} & \multicolumn{2}{l|}{C\&W (one-hot)} \\ \hline
			\multicolumn{1}{|m{1.6cm}|}{Parameters}       & ASR              & PSNR             & ASR            & PSNR           & ASR             & PSNR             \\ \hline
			\multicolumn{1}{|m{1.6cm}|}{(1e-3,10, 100,0)}   & 29.3\%             & 21.26            & 26\%            & 21.19          & 1.5\%            & 32.48       \\ \hline
			\multicolumn{1}{|m{1.6cm}|}{(1e-3,10, 200,0)}   & 43.6\%             & 21.49            & 35.6\%           & 20.73          & 8\%            & 27.69             \\ \hline
			\multicolumn{1}{|m{1.6cm}|}{(1e-3,10, 500,0)}   & 55.6\%             & 21.91            & 43.6\%           & 20.37          & 40.5\%        & 24.29             \\ \hline
			\multicolumn{1}{|m{1.6cm}|}{(1e-3,10, 1000,0)}   & 64.6\%             & 22.23            & 49\%           & 20.56          & 66.5\%        & 24.97          \\ \hline
			\multicolumn{1}{|m{1.6cm}|}{(1e-1,10, 2000,0)} & 72.3\%             & 22.35             & 57.6\%           & 20.35           & 78\%            & 25.29            \\ \hline
		\end{tabular}
		}
\end{table}

\subsection{Methodology}

In \cite{ECOC}, the authors tested the robustness of the ECOC architecture for various combinations of codeword matrices $\mathbf{C}$,  activation functions $\sigma(\cdot)$ and network structures. In particular, they considered the MNIST \cite{MNIST} and CIFAR-10 \cite{cifar10} classification tasks (M = 10 in both cases). In the end, the best performing system was obtained by considering a Hadamard code  with $N=16$ and the tanh activation function. An ensemble of 4 ($N/4$) networks each one outputting 4 bits was considered.
The authors argue that using a large number of ensembles increases the performance of the system against attacks (by decreasing the dependency among output bits).
Then, in our experiments,  we used $N$ ensembles, with only one output bit each.  
\hl{The authors also indicate that the robustness of ECOC scheme can be improved by using longer codewords. Then in our experiments, in addition to MNIST and CIFAR-10 already considered in \cite{ECOC}, we also considered traffic sign classification (GTSRB dataset) \cite{GTSRB}, to test the robustness of ECOC on a larger number of classes and with codewords of a larger size, which potentially means higher robustness. To be specific, for traffic sign classification, we set $M= 32$, by selecting the classes with more examples among the total number of 44 classes in GTSRB, and chose a  Hadamard code with $N =  M = 32$, which is twice the size of the code used for MNIST and CIFAR-10.}
A diagram of the ECOC scheme with the $N$ ensemble structure is shown in Figure \ref{fig:ecoc}.
We used a standard VGG-16 network \cite{VGGnet} as the base block of our implementation. Following the ECOC design scheme, the first 6 layers form the so called {\em shared bottom} part, that is, the layers shared by all the networks of the ensemble. Then, the remaining 10 layers (the last 8 convolutional layers and the 2 fully connected layers) are trained separately for each ensemble branch.

%
For each task, we first trained one  $M$-class classification network, then we fine-tuned the weights to get the $N$ ensemble networks. The error rates of the trained models on clean images are equal to \hl{2.14\% for MNIST,} 13.9\% for CIFAR-10 and 1.28\% for traffic sign (GTSRB) classification.

\hl{ In addition to the extended C\&W attack described in Section \ref{sec:attack}, we also considered \BTT{a new attack named Layerwise Origin-Target Synthesis (LOTS)} introduced in \cite{LOTS}. In a few words, LOTS aims at modifying the deep representation at a chosen attack layer, by minimizing the Euclidian distance between the deep features of the to-be-attacked input, and a target deep representation chosen by the attacker. In our tests, we applied LOTS to the logits level, and we obtained the target deep representation (logits) by randomly  choosing 50 images belonging to the target class. }

\subsection{Results}

We attacked \BTT{300} images randomly chosen from the test set of each task. For each attack, we carried out a targeted attack with the target class chosen at random among the remaining $M-1$ classes (i.e., all the $M$  classes except the original class of the unperturbed image). The label $t$ of the target class was used to run the C\&W attack in eq. \eqref{eq:CWC} and LOTS, while  the codeword $\mathbf{C}_t$ associated to $t$ is considered  in \eqref{eq:CWB} for the new  attack.
We use the Attack Success Rate (ASR) to measure the effectiveness of the attack, i.e. the percentage of generated adversarial examples that are assigned to the target class, and the Peak Signal-to-Noise Ratio (PSNR) to measure the distortion introduce by the attack, which is defined as $\text{PSNR} = 20 * \log_{10}(255 * \sqrt{N}/||\delta||_2)$, where $||\delta||_2$ is the $L_2$ norm of the perturbation and $N$ is the size of the image.

\begin{table*}[!th]
	\caption{Probability values output by the ECOC classifier on CIFAR-10 for different confidence margins of the attack. The parameters of the attacks are indicated according to the following format: (starting point, number of steps of binary search, max iterations, confidence). Prob true and target class indicate the probabilities of the original (true) and target classes, before (B) and after (A) the attack. }
	{

		\label{tab:prob}
		\scalebox{0.8}{	
		\begin{tabular}{|l|l|l|l|l|l|}
			\hline
			{\bf C\&W attack}       & (1e-4,5,500,0)      & (1e-4,5,500,8)     & (1e-4,5,500,12)   & (1e-4,5,500,14)   & (1e-4,5,500,15)    \\ \hline
			ASR      & 62.3\%                & 44.6\%               & 44.6\%              & 42.6\%             & 42.3\%               \\ \hline
			PSNR  (dB)      & 39.94               & 40.78              & 39.57             & 39.00             & 38.40              \\ \hline
			Prob. true class & (B) 0.881  (A) 0.251 & (B) 0.881 (A) 0.153  & (B) 0.881 (A) 0.084 & (B) 0.881 (A) 0.043 & (B) 0.881 (A) 0.021 \\ \hline
			Prob. target class & (B) 0.013 (A) 0.328   & (B) 0.013 (A) 0.534  & (B) 0.013 (A) 0.721 & (B) 0.013 (A) 0.843 & (B) 0.013 -(A) 0.914  \\ \hline
		\end{tabular}
		}
		\\[2pt]
		\scalebox{0.8}{	
		\begin{tabular}{|l|l|l|l|l|l|}
			\hline
			{\bf Proposed attack}   & (1e-4,5,500,0)   & (1e-4,5,500,1.5) & (1e-4,5,500,2.5) & (1e-4,5,500,4.0) & (1e-4,5,500,5.0) \\ \hline
			ASR      & 88.0\%             & 87.6\%             & 86.3\%             & 85.1\%             & 82.7\%             \\ \hline
			PSNR  (dB)      & 38.53            & 37.48            & 37.02            & 36.07             & 35.40            \\ \hline
			Prob. true class & (B) 0.908 (A) 0.194& (B) 0.908 (A) 0.063 & (B) 0.908 (A) 0.024  & (B) 0.908 (A) 0.005  & (B) 0.908 (A)0.001   \\ \hline
			Prob. target class & (B) 0.009 (A) 0.546  & (B) 0.009 (A) 0.824  & (B) 0.009 (A) 0.923  & (B) 0.009 (A) 0.981  & (B) 0.009 (A)0.993   \\ \hline
		\end{tabular}
		}
	}{}
\end{table*}

\hl{As the parameters of the new attack have the same meaning as those of C\&W attack, we first compare the C\&W and the new attack with several settings of the input parameters. The results we got are shown in Table \ref{tab:GTSRB-1}, \ref{tab:Cifar-1} and \ref{tab:MNIST},  for GTSRB, CIFAR-10 and MNIST, respectively. In all the cases, $c$ was set to $0$. The results obtained by using the C\&W attack against the standard one-hot-encoding VGG-16 network with $M$ classes are also reported in the last column.}
\hl{By looking at the different rows, we can first see that when the strength of the attack is increased, e.g. by using a larger number of iterations or a larger number of steps during the binary search, the ASR of both attacks increases, at the price of a slightly larger distortion. \BTT{For instance, for CIFAR-10, the ASR of the proposed attack increases from 69.3\% to 98.6\%, with a decrease in the PSNR of less than 1dB, and the ASR of the C\&W attack increases from 53.6\% to 92.6\% with an extra distortion of 3dB.} Then, by comparing different columns, we see a clear advantage of the proposed attack over C\&W attack, since the former achieves a higher ASR for the same parameter settings.}

\hl{By comparing the different tables, we see that the advantage of the new attack is more evident with GTSRB than with CIFAR-10. The use of longer codewords in GTSRB, in fact, makes it harder to attack this classifier, however the new attack can still achieve an \BTT{ASR = 93.3\%} with a PSNR equal to 39dB}

\hl{For MNIST dataset, the ASR is lower compared to the CIFAR-10 and GTSRB. This result agrees with the results reported in \cite{ECOC}. One possible explanation of this fact is advanced in \cite{EnsembleAdvTraining} where the peculiarities of the MNIST dataset are highlighted and used to argue that high robustness can be easily reached on MNIST.}

\begin{figure}[htp]
\subfloat[GTSRB]{%
  \includegraphics[clip,width=\columnwidth]{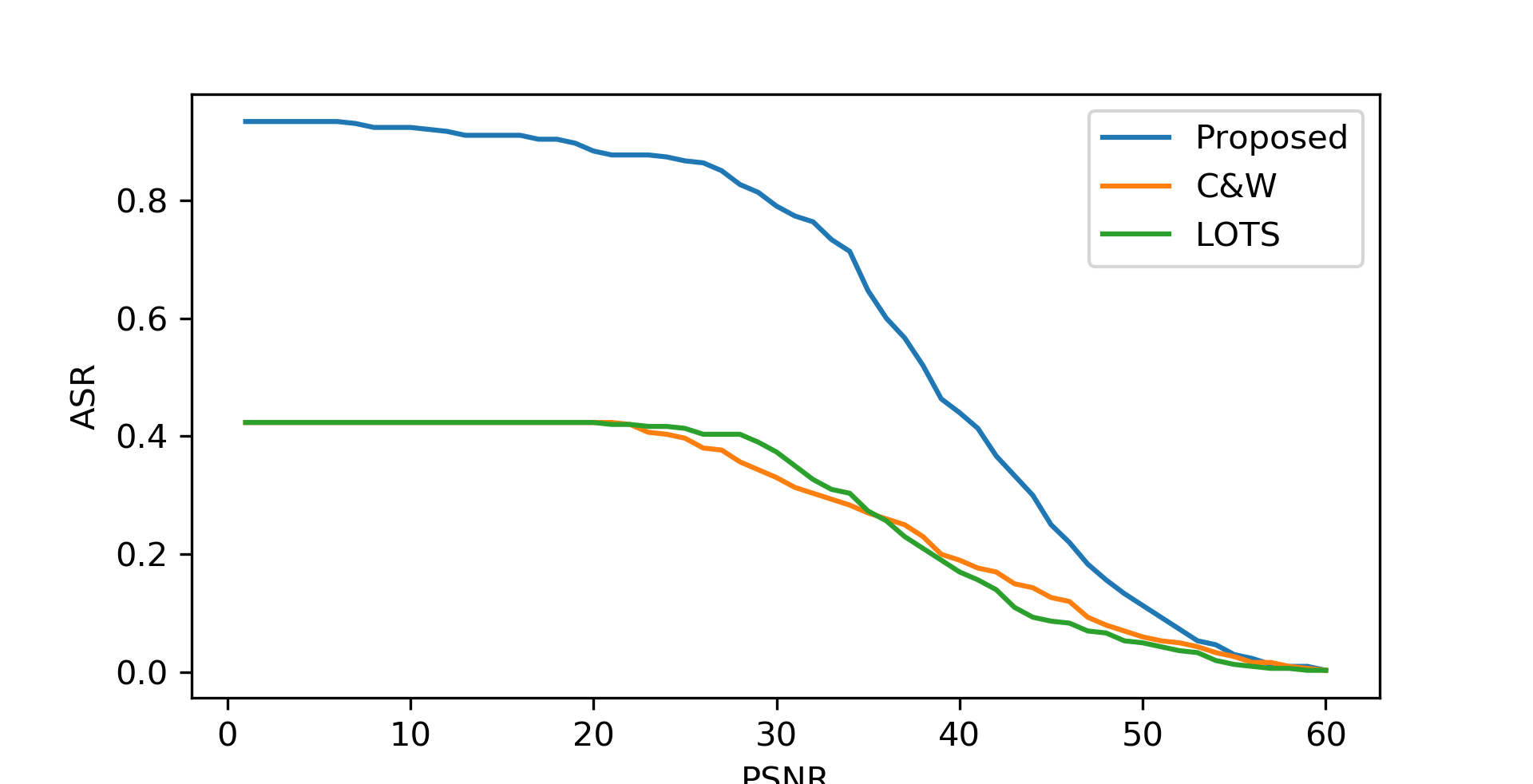}%
}

\subfloat[Cifar]{%
  \includegraphics[clip,width=\columnwidth]{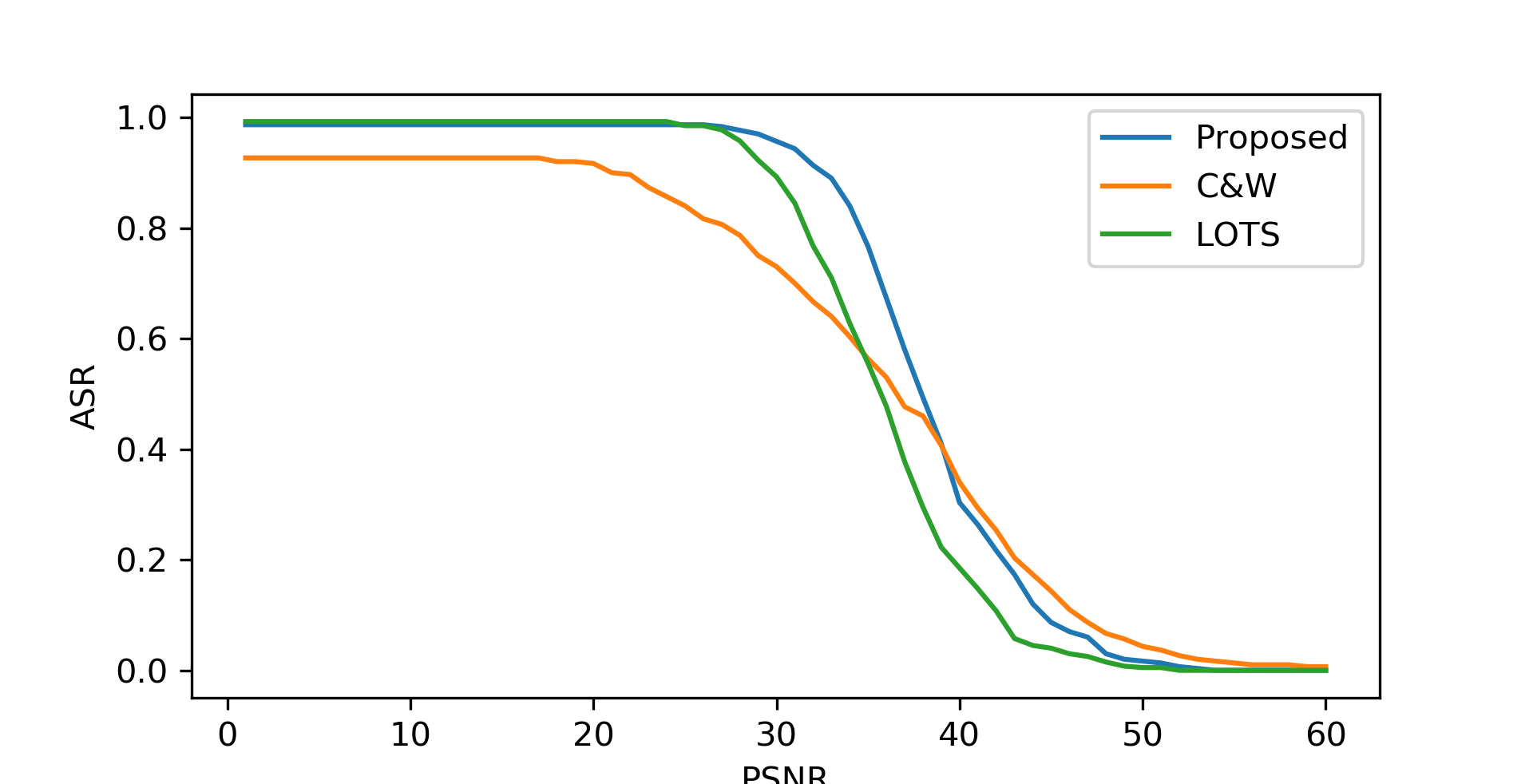}%
}

\subfloat[MNIST]{%
  \includegraphics[clip,width=\columnwidth]{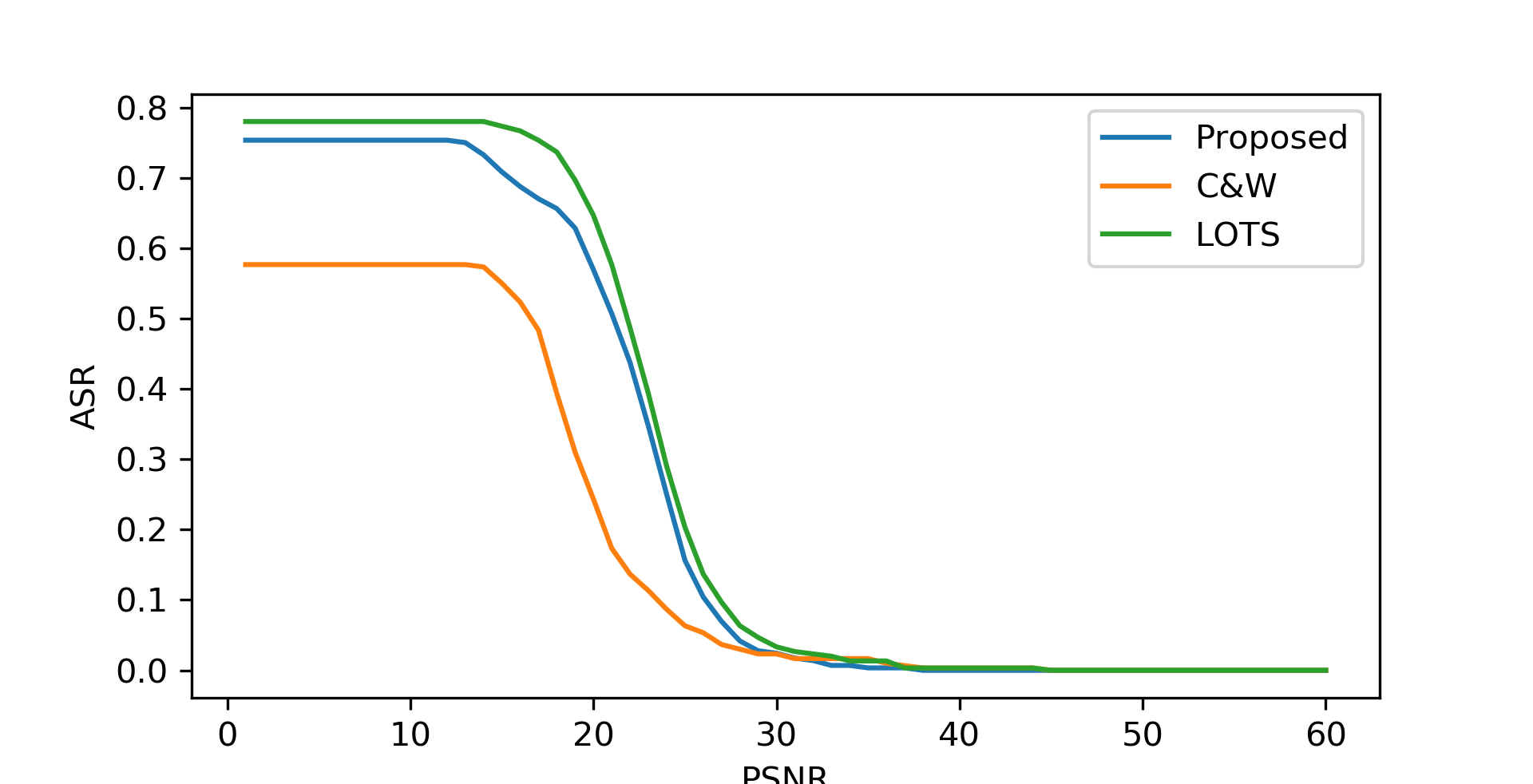}%
}
	
	\caption{Performance of different attacks against the ECOC system, x-axis indicates the PSNR(db) and y-asxis indicates the ASR.}
\label{fig:LOTS}
\end{figure}

\begin{figure}[htb]
\includegraphics[clip,width=\columnwidth]{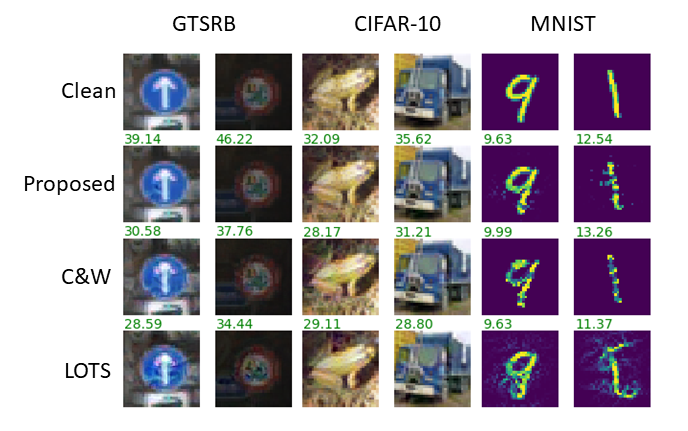}
\caption{Examples of attacked images for different attacks. Besides the first row of clean images, the green number locates at top left of each attacked image indicates its PSNR value.}
\label{fig:img_example}
\end{figure}

\hl{The comparison with LOTS must be carried on a different ground, since such an attack is designed in a different way and the only parameter shared with the new attack is the maximum number of iterations allowed in the gradient descent. For this reason, we applied LOTS by allowing a maximum number of iterations equal to 2000, which is the same number we have used for the other two attacks. We have verified experimentally that LOTS converges within 1000 iterations 92\% of the times\footnote{\scriptsize{\hl{The convergence is determinate by checking whether the new loss value is close enough to the average loss value of the last 10 iterations.}}}, thus validating the adequacy of our choice. Then, we measured the ASR for a given maximum PSNR, thus allowing us to plot the ASR as a function of PSNR.  
The results we got are shown in Figure \ref{fig:LOTS}. Upon inspection of the Figure, we observe a behaviour similar to Table \ref{tab:GTSRB-1}, \ref{tab:Cifar-1} and \ref{tab:MNIST}. The proposed attack greatly outperforms LOTS and C\&W on GTSRB when longer codewords are used. The ASR of the new attack, in fact, achieves nearly 100\% for smaller PSNR's while  LOTS and C\&W stop at \BTT{42\% and 42.3\%}, respectively. For the other two datasets, the gap between the different attacks is smaller than in the GTSRB case. Specifically, the proposed attack and LOTS perform almost the same on CIFAR-10, while LOTS provides slightly better results on MNIST.}
This observation can also be verified in Figure \ref{fig:img_example}, where we show some images that are successfully attacked by all the attacks. We can see from the figure, that the proposed attack requires less distortion to attack the selected examples, producing images that look visually better than the others. The advantage is particularly evident for the GTSRB case, but is still visible for the CIFAR-10 and MNIST images.

\BTT{As for time complexity, we observe that though our attack aims at modifying fewer bits each time, its complexity is very similar to that of C\&W attack. Specifically, if we allow 2000 iterations(10 binary searches) for each attack, for CIFAR-10 the new attack and C\&W attack require about 800 seconds and 1000 seconds to attack an image, respectively\footnote{\BTT{The times are measured using one NVIDIA RTX2080 GPU without paralleling}}. On the other hand, LOTS is considerably faster, since it needs about 80 seconds to attack an image. The reason behind the high computational complexity of C\&W and our new attack, is the binary search carried out at each step. In fact, we verified that by reducing the number of steps the binary search consists of, the speed of both attacks improves greatly. However, since our main purpose is to test the robustness of the ECOC system, we did not pay much effort to optimise our attack from a computational point of view, all the more that its complexity is already similar to that of C\&W attack.}

\hl{As an overall conclusion, the experimental analysis reveals that, in the white-box scenario, the security gain that can be achieved through the ECOC scheme is quite limited, since by properly applying existing attacks and especially by using the newly proposed attack, the ECOC classifiers could be attacked quite easily.}

Another expected advantage of ECOC is that adversarial examples tend to be classified with a lower probability than clean images. Here we show that such a behavior can be inhibited - at the price of a slightly larger distortion - by increasing the confidence of the attack. If a larger confidence margin $c$ is used, in fact, the model becomes more certain about the wrongly predicted class. \hl{To back such a claim, in Table \ref{tab:prob} we report the results of the new attack for different confidence values $c$ for the CIFAR-10 case}\footnote{\scriptsize{\hl{To clearly show the effect of confidence, we did not consider adversarial examples that do not reach the chosen confidence margin $c$, which leads to a slight drop of the ASR.}}}. The table shows the average probability assigned by the ECOC model to the original class (Prob. true class) and to the target class of the attack (Prob. targ class), before and after the attack.

From the table, we see that, by increasing $c$,
\hl{the adversarial examples are assigned higher and higher probabilities for the target class, getting closer to those of the benign samples. In particular, the average probability  for the target class passes from \BTT{0.546} (with $c=0$) to \BTT{0.993} (with $c = 5$), which is even higher than the average probability of the clean images before the attack (0.908), and the probability of the original (true) class decreases from \BTT{0.194} (with $c=0$) to a value lower than 0.001 (with $c = 5$). A similar behavior can be observed for the C\&W attack when $c$ is raised from 0 to 15}. 

\begin{figure}[htp]

	\includegraphics[clip,width=\columnwidth]{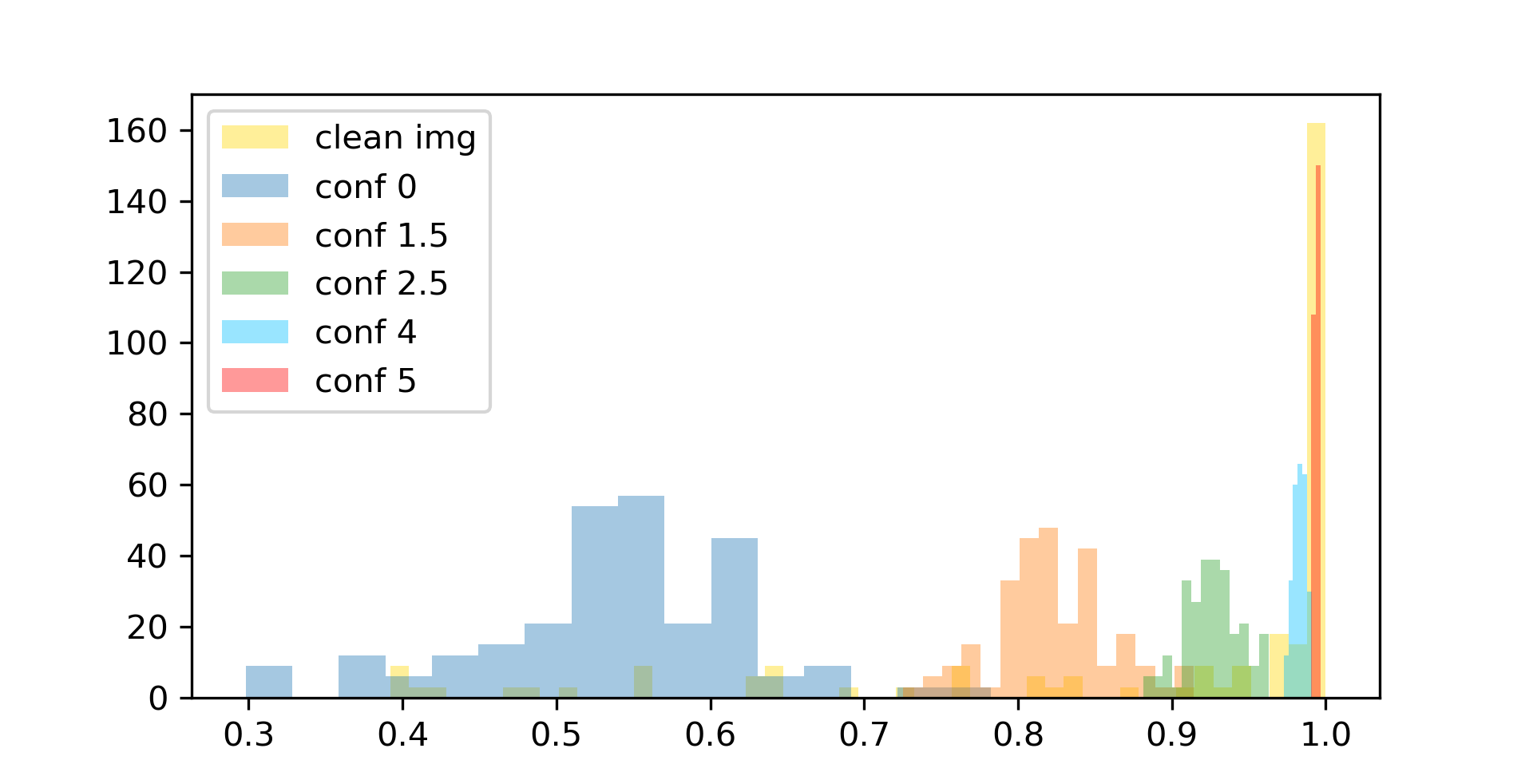}
	
	\caption{Distribution of probabilities assigned to the most probable class of the attacked examples with the proposed attack on CIFAR-10. The $ x $-axis indicates the probability, the $ y $-axis the number of examples classified with such a probability.}
\label{fig:prob dist}	
\end{figure}

Figure \ref{fig:prob dist} shows the distribution of the probabilities assigned to the most probable class for clean and adversarial images generated by the proposed attack. The plot confirms that the ECOC classifier assigns low probabilities only in the presence of adversarial examples obtained with a low confidence value $c$. When $c$ grows, in fact, the probability distribution of adversarial examples get closer and closer to that of clean images, \hl{when $c=5$, it becomes impossible to distinguish clean images and adversarial examples by setting a threshold on the probability assigned to the most probable class.}

\section{Related works}

Adversarial examples, i.e. small, often imperceptible, ad-hoc perturbations of the input images, have been shown to be cable to easily fool neural networks \cite{Intriguing,Explaining and Harnessing,DNNeasyfooled,deepfool}
and have received great attention in the last years.

\hl{Different attacks have been proposed to obtain adversarial examples in various ways. Some works focus on diminishing the computational cost necessary to build the adversarial examples \cite{Explaining and Harnessing, PhysicalFGSM}, while others aim at lowering the perturbation introduced to generate the adversarial examples \cite{Intriguing, CW attack, JSMA}. There are also some works whose goal is to find adversarial examples that modify only one pixel \cite{one pixel attack}, and adversarial perturbations that can either fool several models at the same time \cite{InputDiversity}, or can be applied to several clean images at the same time\cite{UniversalPerturbation}.}

\hl{As a response to the threats posed by the existence of adversarial examples and by the ease with which they can be created, many defence mechanisms have also been proposed. According to \cite{WildPatterns} and \cite{SuveryTNNLS2019}, defences can be roughly categorized into two branches, that either work in a reactive or proactive manner. The first class of defences is applied after that the DNNs have been built. This class includes approaches exploiting randomization, like for instance stochastic activation pruning, in which node activations at each (or some) layers are randomly dropped out during the forward propagation pass \cite{pruning}, and, more recently, model switching \cite{Switching}, where random selection is performed between several trained sub-models. Other approaches attempt to intentionally modify the network input to mitigate the adversarial perturbation, e.g. by projecting the input into a different space \cite{inputdenfens1} or by applying some input transformations \cite{inputdefense2}. Other approaches attempt to reject input samples that exhibit an outlying behavior with respect to the unperturbed training data \cite{interpretability}. The second branch of defences aims at building more robust DNNs. One simple approach to improve the robustness against adversarial example is adversarial
training, which consists in augmenting the training set with adversarial examples \cite{advtraining1,advtraining2,rPDA,EnsembleAdvTraining}. More recently, as more attention has been paid to hidden layers with respect to the robustness of DNNs \cite{Cisse2017}, it has been proven that rather than augmenting the training set, the robustness of DNNs can be improved by directly injecting adversarial noise into the hidden nodes, thus improving the robustness of single neurons \cite{rANP,rNeuronSensity} }

\hl{The ECOC scheme considered in this paper, belongs to the second class of defences and is derived from similar attempts made in the general machine learning literature  to improve the robustness of multi-class classification problems \cite{ecoc1,eoco2,eoco3}}.
The robustness of ECOC against adversarial examples is assessed in \cite{ECOC} by considering conventional adversarial attacks like \cite{PGD,CW attack}, which have not been explicitly designed for multi-label classification. As suggested in \cite{AdaptiveAttack}, however, in order to properly assess the effectiveness of a defence
mechanism, the case of an informed attacker should be considered, and then the robustness should be evaluated against attacks targeting the specific defence mechanism.
Following the spirit of \cite{AdaptiveAttack}, in this paper we developed a targeted attack against the ECOC system, that exploits the multi-class and multi-label nature of such a system. \hl{We observe that the capability of ECOC to hinder the generation of adversarial examples has already been challenged in \cite{AdaptiveAttack} (Section10), however the analysis in \cite{AdaptiveAttack} is carried out under the more favourable (for the attacker) assumption of a non-targeted attack, thus marking a significant difference with respect to the current work.} 
\section{Conclusion}
In order to investigate the effectiveness of ECOC-based deep learning architectures to hinder the generation of adversarial examples, we have proposed a new targeted attack explicitly thought to work with such architectures. We measured the validity of the new attack experimentally on \hl{three common classification tasks, namely  GTSRB, CIFAR-10 and MNIST}. The results we have got show the effectiveness of the new attack and, most importantly, demonstrate that the use of error correction to code the output of a CNN classifier does not increase significantly the robustness against adversarial examples, \hl{even in the more challenging case of a targeted attack. In fact, the ECOC scheme can be fooled by introducing a small perturbation into the images, both with the new attack and, to a lesser extent, by applying C\&W  and LOTS attacks with a proper setting}. No significant advantage in terms of decision confidence is observed as well, given that by properly setting the parameters of the attack, adversarial examples are assigned to the wrong class with a high probability.

\section{Data Availability}
The data used to support the findings of this study are available from the first author(bowenzhang.psnl@outlook.com) upon request. 

\section{Conflicts of Interest}
The authors declare that there are no conflicts of interest regarding the publication of this paper.

\section*{Funding Statement}
This work has been partially supported by the 	China Scholarship Council(CSC), file No.201806960079.

\bibliographystyle{plain}

\end{document}